\documentclass[prc,preprint]{revtex4}

\usepackage{graphicx}
\usepackage{amsmath}
\usepackage{color}
\begin{document}

\title{{\bf Superluminal light propagation in a normal dispersive medium}}

\author{{\bf Zahra Amini Sabegh and Mohammad Mahmoudi \footnote{E-mail: mahmoudi@znu.ac.ir}}}
 \affiliation{Department of Physics, University of Zanjan, University Blvd., 45371-38791, Zanjan, Iran}

\begin{abstract}

We study the propagation of a Laguerre-Gaussian (LG) beam through a dispersive atomic medium. We restrict ourselves to applying a weak probe field and three strong coupling fields to the medium, which leads to developing a four-level double $V$-type atomic system. We first regard all the three strong coupling fields as the plane-waves and calculate an analytical expression for the group velocity of the probe LG field on the optical axis at the waist of the field. It appears that the resulting formula in a dispersive medium is in good agreement with that of the free space. We also find a more general analytical expression for the group velocity of the probe LG field out of the optical axis and compare with its projection onto the propagation axis. It is turned out that these two quantities are equal on the optical axis, at the waist of the beam and the Rayleigh range. Finally, we assume one of the strong coupling fields to be an LG field and explore how its orbital angular momentum (OAM) affects the group velocity of the probe LG field. Our analysis predicts a strange behavior for the group velocity of the probe LG field inside a normal dispersive medium so that it can exceed the speed of light in free space. Such an unusual propagation of the LG light beam results from the distortion of its helical phase front via the classical interference of the planar and LG fields.

\end{abstract}
\maketitle

\section{Introduction}

A wide variety of theoretical and experimental studies have recently been conducted on the group velocity of structured light fields propagating in free space. Padgett \textit{et al.} have experimentally explored how the transverse spatial structure of photons affects the dispersion of light beam in free space. It was observed that the group velocity of structured single photons is slower in both Bessel and focused Gaussian beams along the propagation axis than the speed of light in free space $c$ \cite{Padgett2015}. In another experimental work, it was demonstrated that the addition of orbital angular momentum (OAM) of the structured beam reduces the intrinsic delay of twisted photons with respect to the same beam without OAM \cite{Howfast2018}. It is noted that the group velocity reduction for the Bessel light beam near the critical frequency has been reported in Ref. \cite{Alfano2016}. On the other hand, the group velocity of the Laguerre-Gaussian (LG) beam with helical wavefront has been theoretically studied and shown that it is inversely proportional to the OAM value of light \cite{Srep2016}. The authors declared that the slowing of light is due to the field confinement and the dispersion of light, which implies its intrinsic property in free space. Theoretical and experimental investigation of the evolution of the group velocity was performed for the focused Gaussian and LG beams along the propagation axis in Ref. \cite{Optica2016}. It was presented that the group velocity of a Gaussian mode can be both subluminal and superluminal in different propagation distances on the optical axis, while an LG mode has the slow-light behavior for all propagation distances. However, a comment \cite{Saari2017} found the results of Ref. \cite{Optica2016} questionable in some aspects and suggested using the projection of group velocity vector onto the beam axis instead of its absolute value, particularly when one interprets the measured relatively large propagation delays. More recently, Saari resolved the contradictions between various definitions of group velocity, which led to the different versions of group velocity for Bessel-Gauss pulses depending on the type of pulse and method of recording it in the output plane \cite{Saari2018}.

Over the last three decades, the propagation of the plane-wave light beam through a dispersive medium has attracted much attention. It is well-known that the group velocity of such beams, interacting with matter, can be either subluminal \cite{Behroozi1999,Kash1999} or superluminal \cite{Steinberg1994,Wang2000}. Generally, the subluminal (superluminal) phenomenon occurs when the medium dispersion has normal (anomalous) behavior \cite{Ficek,Scully}. In recent years, both of these phenomena and also switching from one to another have been studied in different quantum systems such as atomic media \cite{Prof1,Prof2,Prof3,Prof4,Prof5,Prof6,Prof2009,Prof7,Prof8}, quantum dot molecules \cite{ProfPhysE2009}, photonic crystals \cite{Sahrai2010}, superconducting quantum circuits \cite{Amini2015}, graphene \cite{Kazemi2018}, and nitrogen-vacancy centers \cite{Ghaderi2019}. It should be noted that the superluminal or subluminal group velocity of a light beam results from its interaction with matter or the structure of the wavefront. Recently, in studying the OAM transfer, we have reported the preliminary results of the group velocity of the probe LG field in a dispersive medium \cite{Mohammad}.

In this manuscript, the group velocity of the LG beam is studied considering two scenarios: uniformly and spatially dependent dispersive medium, in which the obtained analytical expressions help us to explore the behavior of the group velocity and its relation with the medium refractive index. The first scenario involves the subluminal group velocity accompanied by normal dispersion under the multi-photon resonance condition. Our analytical expression, which is calculated for the group velocity of the probe LG field in the dispersive atomic medium, can reduce to its well-known formula in free space. Moreover, we investigate the group velocity and its $z$-component along the propagation direction (out of the optical axis) and demonstrate that there are some regions where both of these quantities are equal employing two mentioned relative phases. On the other hand, the effect of the OAM of strong coupling LG field on the group velocity of the probe LG field is studied through the second scenario. It is shown that the superluminal (subluminal) group velocity with normal (anomalous) dispersion appears in some regions of the medium. Such a strange result is the effect of the classical interference of the coupling planar and LG fields on the helical phase front of the probe LG field, which has not yet been reported before in a dispersive medium.

\section{Results and discussion}

\subsection{Theoretical model}
\begin{figure}[htbp]
\centering
  \includegraphics[width=0.75\linewidth]{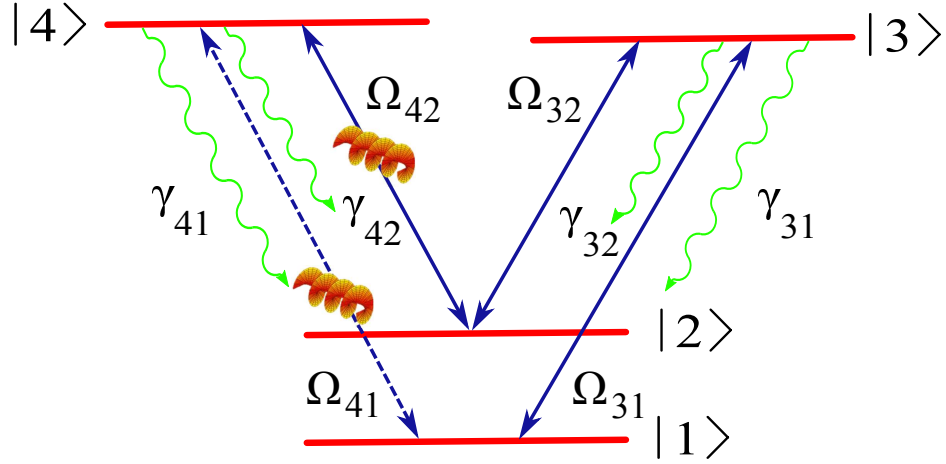}
  \caption{\small Schematic diagram of the four-level double $V$-type atomic system driven by a weak probe field ($\Omega_{41}$) and three strong coupling fields ($\Omega_{31}$, $\Omega_{32}$, and $\Omega_{42}$).}\label{fig1}
\end{figure}

Let us consider an atomic vapor cell consisting of $^{87}Rb$ atoms, whose four selected transitions related to its $D_1$ line are driven by four external fields. As shown in Fig. \ref{fig1}, we establish a four-level double $V$-type atomic system in the suggested model,
and choose the four energy states from the $D_1$ line of $^{87}Rb$ atoms: $|1\rangle=|5~^2S_{1/2},F=1\rangle$, $|2\rangle=|5~^2S_{1/2},F=2\rangle$, $|3\rangle=|5~^2P_{1/2},F=1\rangle$, and $|4\rangle=|5~^2P_{1/2},F=2\rangle$. In our notation, the Rabi frequency $\Omega_{ij}=\vec{\mu}_{ij}\cdot\vec{E}_{ij}/\hbar$ is defined as the scale of the applied field strength $E_{ij}$ in which $\mu_{ij}$ and $\hbar$ are the induced dipole moment of the transition $|i\rangle\leftrightarrow|j\rangle$ and Planck's constant, respectively. It is assumed that the $|1\rangle\leftrightarrow|4\rangle$ transition is excited by a weak probe LG field with the probe detuning $\Delta_{41}$, and the Rabi frequency in cylindrical coordinates
\begin{eqnarray}\label{e1}
    \Omega_{41}(r,\varphi,z)=\Omega_{41_0}(\frac{\sqrt{2}r}{w_0})^{|l_{41}|}e^{-\frac{r^2}{w_0^2}}L_{p_{41}}^{|l_{41}|}(2r^2/w_0^2)e^{i\Phi(r,\varphi,z)}.
\end{eqnarray}
Here $L_{p_{41}}^{|l_{41}|}$, $l_{41}$, $p_{41}$, $w_0$, and $\Omega_{41_0}$ stand for the associated Laguerre polynomial, OAM value, radial index, waist, and constant Rabi frequency of the LG field, respectively. The phase of the LG field described by $\Phi(r,\varphi,z)$ has the following form
\begin{eqnarray}\label{e2}
    \Phi(r,\varphi,z)=\frac{n\omega}{c}z(1+\frac{r^2}{2(z^2+z_R^2)})
    -(2p_{41}+|l_{41}|+1)tan^{-1}(z/z_R)+l_{41}\varphi,
\end{eqnarray}
where the refractive index of the medium, laser frequency, and the Rayleigh range are indicated by $n$, $\omega$, and $z_R=n\omega w_0^2/2c$, respectively. Three strong coupling fields are also applied to three transitions of the atomic system, $|1\rangle\leftrightarrow|3\rangle$, $|2\rangle\leftrightarrow|3\rangle$, and $|2\rangle\leftrightarrow|4\rangle$, with the Rabi frequencies ($\Omega_{31}$, $\Omega_{32}$, and $\Omega_{42}$) and frequencies ($\omega_{31}$, $\omega_{32}$, and $\omega_{42}$) included. Furthermore, spontaneous emission rates from the upper energy states, $|3\rangle$ and $|4\rangle$, to the lower states, $|1\rangle$ and $|2\rangle$, are denoted by $\gamma_{31}$, $\gamma_{32}$, $\gamma_{41}$, and $\gamma_{42}$.

In order to describe the interaction of the applied fields with the considered atomic system, we can use the von Neumann equation for the density matrix of the system, $i\hbar\partial\rho/\partial t=[H,\rho]$. Considering the electric-dipole moment and rotating-wave approximations in the interaction picture, the Hamiltonian can be written as
\begin{eqnarray}\label{e3}
    H=-\hbar[\Omega_{31}^\ast e^{i\Delta_{31}t}|1\rangle\langle3|+\Omega_{41}^\ast e^{i\Delta_{41}t}|1\rangle\langle4|
    +\Omega_{32}^\ast e^{i\Delta_{32}t}|2\rangle\langle3|
    +\Omega_{42}^\ast e^{i(\Delta_{42}t-\phi_0)}|2\rangle\langle4|+C.C.],
\end{eqnarray}
where $\phi_0$ and $\Delta_{ij}=\omega_{ij}-\bar{\omega}_{ij}$ are the relative phase of the applied fields and frequency detuning between laser frequency ($\omega_{ij}$) and central frequency of the corresponding transition ($\bar{\omega}_{ij}$), respectively. The Bloch equations are then obtained for the density matrix elements by substituting Eq. (\ref{e3}) into the von Neumann equation
\allowdisplaybreaks
\begin{eqnarray}\label{e4}
\dot{\rho}_{11}&=&i[\Omega_{31}^{*}\rho_{31}-\Omega_{31}\rho_{13}+\Omega_{41}^{*}\rho_{41}-\Omega_{41}\rho_{14}]
+\gamma_{31}\rho_{33}+\gamma_{41}\rho_{44},\nonumber\\
\dot{\rho}_{22}&=&i[\Omega_{32}^{*}\rho_{32}-\Omega_{32}\rho_{23}+\Omega_{42}^{*}e^{-i\phi_0}\rho_{42}-\Omega_{42}e^{i\phi_0}\rho_{24}]
\gamma_{32}\rho_{33}+\gamma_{42}\rho_{44},\nonumber\\
\dot{\rho}_{33}&=&i[\Omega_{31}\rho_{13}-\Omega_{31}^{*}\rho_{31}+\Omega_{32}\rho_{23}-\Omega_{32}^{*}\rho_{32}]
-(\gamma_{31}+\gamma_{32})\rho_{33},\nonumber\\
\dot{\rho}_{12}&=&i[\Omega_{31}^{*}\rho_{32}-\Omega_{32}\rho_{13}+\Omega_{41}^{*}\rho_{42}-\Omega_{42}e^{i\phi_0}\rho_{14}
-(\Delta_{31}-\Delta_{32})\rho_{12}],\nonumber\\
\dot{\rho}_{13}&=&i[\Omega_{31}^{*}(\rho_{33}-\rho_{11})+\Omega_{41}^{*}\rho_{43}-\Omega_{32}^{*}\rho_{12}-\Delta_{31}\rho_{13}]
-\frac{(\gamma_{31}+\gamma_{32})}{2}\rho_{13},\nonumber\\
\dot{\rho}_{14}&=&i[\Omega_{41}^{*}(\rho_{44}-\rho_{11})+\Omega_{31}^{*}\rho_{34}-\Omega_{42}^{*}e^{-i\phi_0}\rho_{12}-\Delta_{41}\rho_{14}]
-\frac{(\gamma_{41}+\gamma_{42})}{2}\rho_{14},\nonumber\\
\dot{\rho}_{23}&=&i[\Omega_{32}^{*}(\rho_{33}-\rho_{22})+\Omega_{42}^{*}e^{-i\phi_0}\rho_{43}-\Omega_{31}^{*}\rho_{21}-\Delta_{32}\rho_{23}]
-\frac{(\gamma_{31}+\gamma_{32})}{2}\rho_{23},\nonumber\\
\dot{\rho}_{24}&=&i[\Omega_{42}^{*}e^{-i\phi_0}(\rho_{44}-\rho_{22})+\Omega_{32}^{*}\rho_{34}-\Omega_{41}^{*}\rho_{21}-\Delta_{42}\rho_{24}]
-\frac{(\gamma_{41}+\gamma_{42})}{2}\rho_{24},\nonumber\\
\dot{\rho}_{34}&=&i[\Omega_{31}\rho_{14}-\Omega_{41}^{*}\rho_{31}+\Omega_{32}\rho_{24}-\Omega_{42}^{*}e^{-i\phi_0}\rho_{32}
-(\Delta_{41}-\Delta_{31})\rho_{34}]-\frac{(\gamma_{31}+\gamma_{32}+\gamma_{41}+\gamma_{42})}{2}\rho_{34},\nonumber\\
\dot{\rho}_{44}&=&-(\dot{\rho}_{11}+\dot{\rho}_{22}+\dot{\rho}_{33}).
\end{eqnarray}
Note that the terms of the spontaneous emissions have been phenomenologically added to the Bloch equations.

We analytically solve Eq. (\ref{e4}) considering the steady-state condition and assumption of $\Delta_{31}=\Delta_{32}=\Delta_{42}=0$ and $\gamma_{31}=\gamma_{32}=\gamma_{41}=\gamma_{42}=\gamma$, for better understanding how the atomic medium reacts to the probe field. The coherence term of the $|1\rangle\leftrightarrow|4\rangle$ transition $\rho_{41}$ corresponding to the third-order susceptibility of the system determines the response of the atomic system to the probe field. Since the analytical expression of the susceptibility is too long, we present the susceptibility and its derivation with respect to the probe detuning, taking $\Delta_{41}=0$ for simplicity as below
\begin{eqnarray}\label{e5}
\chi=\frac{-i\gamma\Omega_{31}\Omega_{32}^\ast\Omega_{42}e^{i\phi_0}}{A_1},
\end{eqnarray}
and
\begin{eqnarray}\label{e6}
\frac{\partial\chi}{\partial\omega}=\frac{\Omega_{31}\Omega_{32}^\ast\Omega_{42}e^{i\phi_0}A_2}{A_1 A_3},
\end{eqnarray}
in which
\begin{eqnarray*}
A_1&=&\gamma^2(|\Omega_{31}|^2+|\Omega_{32}|^2+|\Omega_{42}|^2)+4|\Omega_{31}|^2|\Omega_{42}|^2,\nonumber\\
A_2&=&2\gamma^4+3\gamma^2(|\Omega_{32}|^2+|\Omega_{42}|^2)+(|\Omega_{32}|^2+|\Omega_{42}|^2)^2
+|\Omega_{31}|^2(\gamma^2+|\Omega_{32}|^2-|\Omega_{42}|^2),\nonumber\\
A_3&=&2\gamma^2(|\Omega_{31}|^2+|\Omega_{32}|^2+|\Omega_{42}|^2)+(|\Omega_{32}|^2+|\Omega_{42}|^2)^2
+|\Omega_{31}|^2(|\Omega_{31}|^2+2|\Omega_{32}|^2-2|\Omega_{42}|^2).
\end{eqnarray*}
It should be noted that the imaginary (real) part of the susceptibility describes the absorption (dispersion) behavior of the atomic system toward the probe field. Contractually, the positive (negative) value for the imaginary part of the susceptibility is assigned to the absorption (gain) of the medium. However, the positive (negative) value of the dispersion slope defines the normal (anomalous) dispersion for the probe field.

The susceptibility of the atomic medium helps to obtain the group velocity through the calculation of the refractive index $n\approx1+Re(\chi)/2$ and dispersion slope $\partial n/\partial\omega$. In the following, we analytically study the group velocity behavior of the probe LG field in three different situations.

\subsection{The group velocity of the probe LG field in a dispersive atomic medium}

The group velocity of the probe LG field can be obtained employing the well-known formula for the magnitude of the group velocity vector, $v_g=1/|\nabla\partial_\omega\Phi|$, where $\partial_\omega\Phi$ is the phase profile derivation of beam, $\Phi$, with respect to $\omega$ \cite{Wolf}. Using Eq. (\ref{e2}) one can calculate the group velocity of the probe LG field in the dispersive medium driven by three strong coupling planar fields, through the propagation axis, $r=0$, and at $z=0$ as below
\begin{eqnarray}\label{e7}
    v_g=\frac{c}{(n+\omega\partial n/\partial\omega)[1+\frac{2c^2}{n^2\omega^2w_0^2}(2p_{41}+|l_{41}|+1)]}.
\end{eqnarray}
Under the mentioned conditions, the vector of group velocity related to the probe LG field has just the z-component, which is given by $v_z=\partial_z\partial_\omega\Phi/|\nabla\partial_\omega\Phi|^2$ \cite{Saari2018}. Note that if $w_0\rightarrow\infty$, the above equation turns into the well-known formula of the group velocity $v_g=c/(n+\omega\partial n/\partial\omega)$ related to a group of plane waves in a dispersive medium \cite{Ficek}. Moreover, it can be deduced from Eq. (\ref{e7}) that the group velocity of the probe Gaussian field ($l_{41}=0$ and $p_{41}=0$) is smaller than $c$ in an atomic medium with normal dispersion. However, its value undergoes a great reduction for the higher modes of probe LG field. However, the probe Gaussian field propagates with a superluminal group velocity when anomalous dispersion dominates the atomic medium. It is also expected from this equation that the higher-order LG modes experience a smaller increase in the group velocity. It is worth noting that our analytical expression for the group velocity of the LG beam in a dispersive medium is in good agreement with the group velocity expression of the LG beam in free space, $n=1$ and $\partial n/\partial\omega=0$, \cite{Srep2016}
\begin{eqnarray}\label{e8}
    v_g=\frac{c}{1+\frac{2c^2}{\omega^2w_0^2}(2p_{41}+|l_{41}|+1)},
\end{eqnarray}
which is smaller than $c$ for a Gaussian beam and all LG modes.
\begin{figure}[htbp]
\centering
  \includegraphics[width=0.49\linewidth]{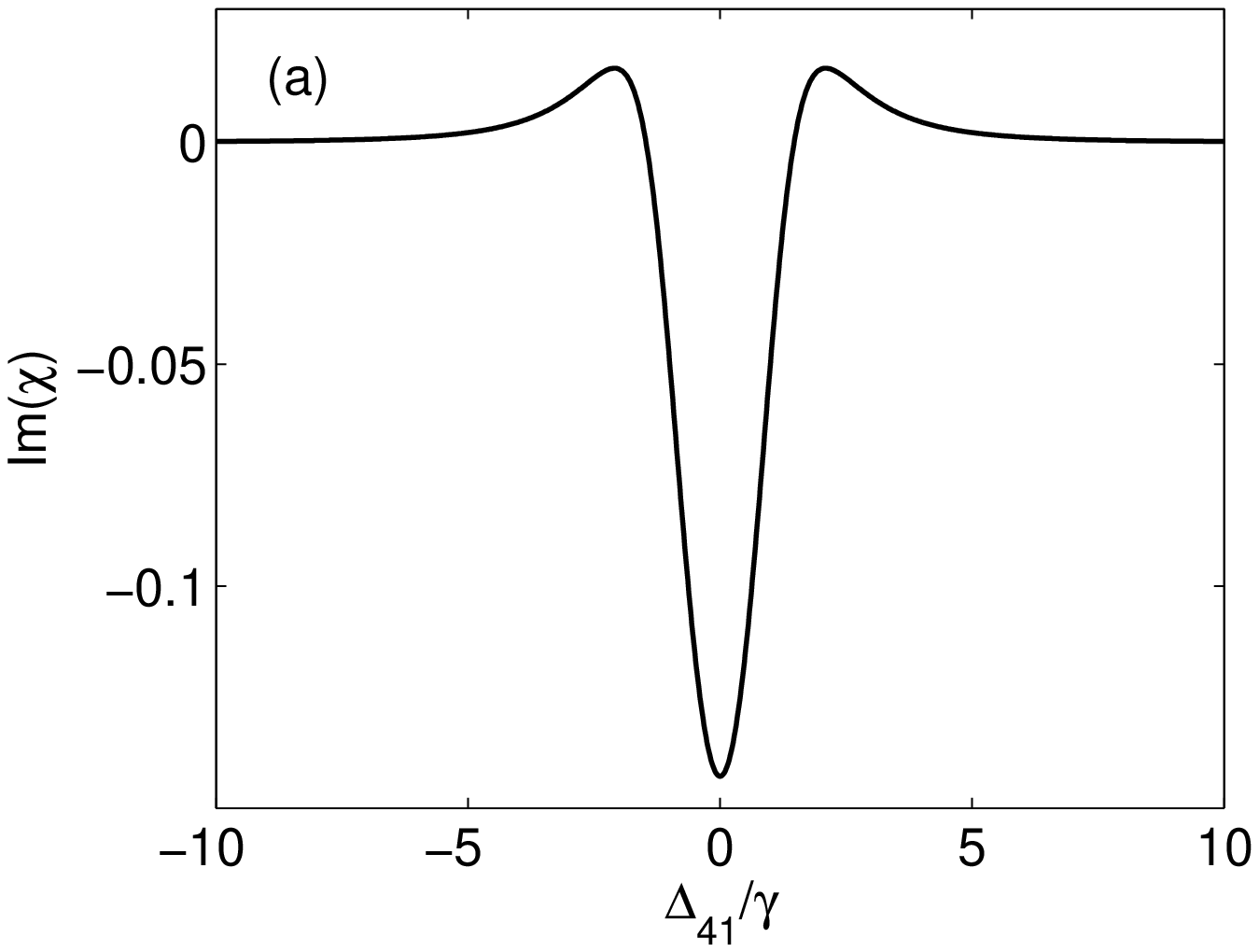} \includegraphics[width=0.49\linewidth]{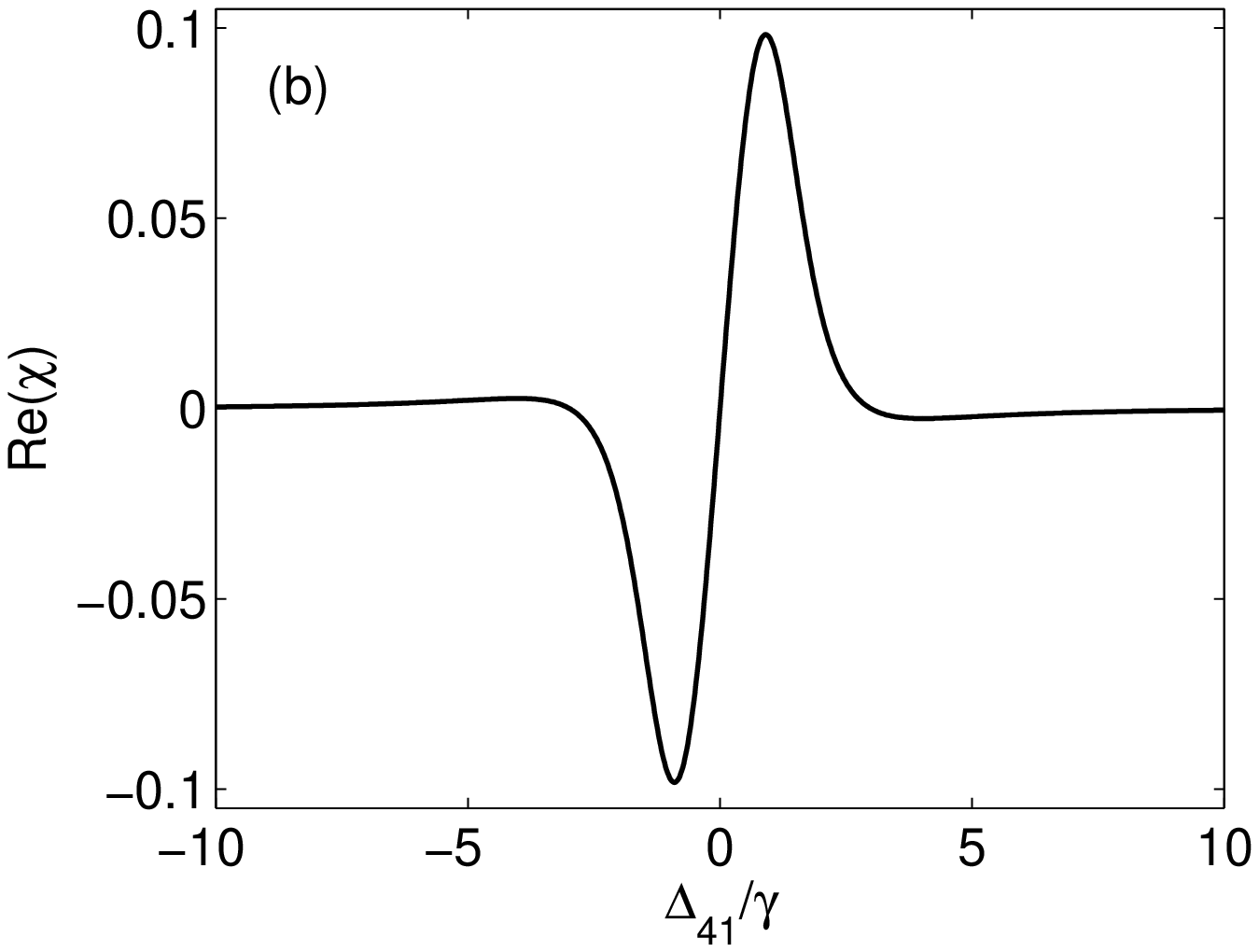}
  \caption{\small The imaginary (a) and real (b) parts of the susceptibility, $\chi$, versus the dimensionless detuning of the probe field, $\Delta_{41}/\gamma$. The applied parameters are considered to be $\Omega_{31}=\Omega_{32}=\Omega_{42}=\gamma$, $\gamma_{31}=\gamma_{32}=\gamma_{41}=\gamma_{42}=\gamma$, $\Delta_{31}=\Delta_{32}=\Delta_{42}=0$, $\Omega_{41}=0.01\gamma$, and $\phi_0=0$.}\label{fig2}
\end{figure}

At this stage, one can study the behavior of the group velocity, medium absorption and dispersion by the obtained analytical results. The imaginary (a) and real (b) parts of the susceptibility, $\chi$, are numerically plotted as a function of the dimensionless detuning of the probe field, $\Delta_{41}/\gamma$, in Fig. \ref{fig2}. The values of parameters are considered to be $\Omega_{31}=\Omega_{32}=\Omega_{42}=\gamma$, $\gamma_{31}=\gamma_{32}=\gamma_{41}=\gamma_{42}=\gamma$, $\Delta_{31}=\Delta_{32}=\Delta_{42}=0$, $\Omega_{41}=0.01\gamma$, and $\phi_0=0$. As exhibited in Fig. \ref{fig2}, a gain dip is accompanied by a normal dispersion around zero probe detuning $\Delta_{41}=0$. Moreover, it is to be noted that the results are consistent with Kramers-Kronig relations.
\begin{figure}[htbp]
\centering
  \includegraphics[width=0.5\linewidth]{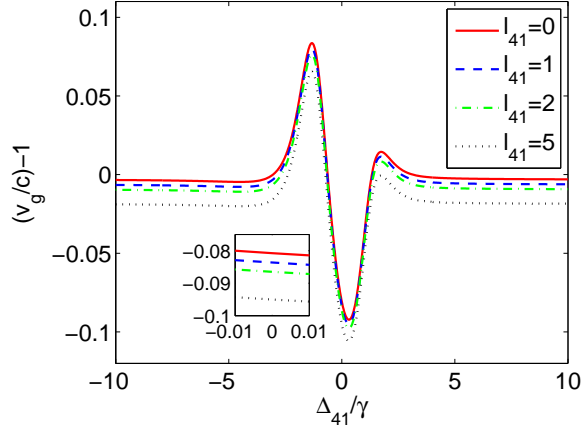}
  \caption{\small The behavior of group velocity, $(v_g/c)-1$, as a function of $\Delta_{41}/\gamma$ for the different modes of probe LG field with $\omega=377.107~THz$, $p_{41}=0$, and $w_0=20\mu m$. Other used parameters are the same as in Fig. \ref{fig2}.}\label{fig3}
\end{figure}

On the other hand, Fig. \ref{fig3} presents the behavior of group velocity, $(v_g/c)-1$, versus $\Delta_{41}/\gamma$ for the different values of OAM, $l_{41}=0$ (solid), $1$ (dashed), $2$ (dash-dotted), $5$ (dotted). The frequency of probe transition, $\omega$, for the $D_1$ line of $^{87}Rb$ atoms is about $377.107~THz$. It should also be mentioned that the values of the radial index and beam waist of probe LG field are taken as $p_{41}=0$ and $w_0=20\mu m$, respectively, while other parameters are the same as in Fig. \ref{fig2}. It is obvious that the group velocity is smaller than $c$ with the multi-photon resonance condition, which is called subluminal light propagation. As deduced from Eq. (\ref{e7}), the probe Gaussian field has a subluminal group velocity when the dispersion slope is positive. Moreover, the probe LG field including higher values of the OAM propagates along $z$-direction slower than the Gaussian mode. Since there are small variations among group velocities of different modes of the probe LG field, a magnified view around zero probe detuning is provided in the inset of Fig. \ref{fig3}.

\subsection{Off-axis group velocity of the probe LG field along propagation direction}

In order to study the behavior of the group velocity out of the optical axis in a dispersive medium excited by three strong coupling planar fields, we recalculate Eq. (\ref{e7}) for $r\neq0$ and $z\neq0$. Therefore, a general expression for the group velocity magnitude takes the following form:
\begin{eqnarray}\label{e9}
    v_g(r,z)=\frac{cB_1^3}{(n+\omega\partial n/\partial\omega)\sqrt{B_2^2+(B_3+B_4-B_5)^2}},
\end{eqnarray}
in which
\begin{eqnarray*}
B_1&=&4c^2z^2+n^2\omega^2w_0^4,\nonumber\\
B_2&=&4c^2rz(-16c^4z^4+n^4\omega^4w_0^8),\nonumber\\
B_3&=&48c^4n^2\omega^2w_0^4z^2(r^2+z^2)+n^6\omega^6w_0^{12},\nonumber\\
B_4&=&2c^2n^4\omega^4w_0^8[-r^2+6z^2+w_0^2(2p_{41}+|l_{41}|+1)],\nonumber\\
B_5&=&32c^6z^4[r^2-2z^2+w_0^2(2p_{41}+|l_{41}|+1)].
\end{eqnarray*}
It can be observed that Eq. (\ref{e9}) reduces to Eq. (\ref{e7}) at $r=z=0$. Here the group velocity vector has a radial component in addition to the $z$-component and thus its projection onto the $z$-axis is not equal to the group velocity magnitude, which is given by
\begin{eqnarray}\label{e10}
    v_z(r,z)=\frac{\partial_z\partial_\omega\Phi}{|\nabla\partial_\omega\Phi|^2}=
    \frac{B_3+B_4-B_5}{\sqrt{B_2^2+(B_3+B_4-B_5)^2}}v_g(r,z).
\end{eqnarray}
The coefficient $(B_3+B_4-B_5)/\sqrt{B_2^2+(B_3+B_4-B_5)^2}$ in Eq. (\ref{e10}) is exactly equal to one at $r=0$, $z=0$, and $z=z_R$. It means that the group velocity vector has just the $z$-component on the optical axis ($r=0$), at the waist of the beam ($z=0$) and the Rayleigh range ($z=z_R$). However, the mentioned coefficient takes the value of about one at other points leading to a difference between $v_z(r,z)$ and $v_g(r,z)$ in this case. Our next step is to study the effect of the OAM value and the relative phase of the applied fields on the radial changes of the group velocity of probe LG field propagating through the atomic medium via Eqs. (\ref{e5}), (\ref{e6}), (\ref{e9}), and (\ref{e10}).
\begin{figure}[h!]
\centering
  \includegraphics[width=\linewidth]{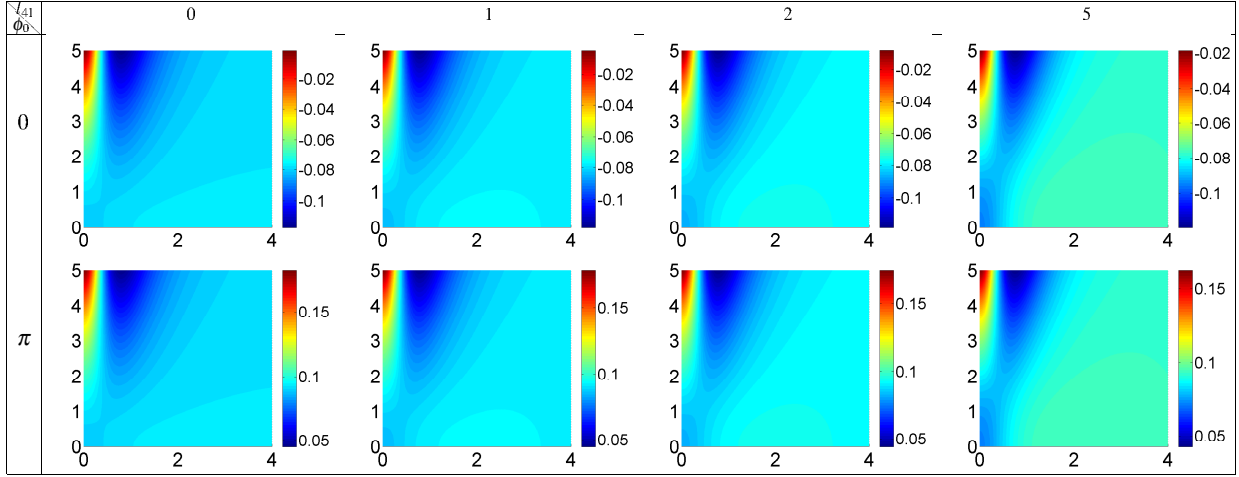}
  \caption{\small The dimensionless group velocity profile of the probe LG field versus the dimensionless propagation distance $z/z_R$ (horizontal axis) and the radial coordinate $r/w_0$ (vertical axis) considering different modes of the probe LG field, i.e. $l_{41}=0,1,2,5$, together with two different relative phases: $\phi_0=0$ (first row) and $\pi$ (second row). All panels are obtained for zero probe detuning $\Delta_{41}=0$, whereas other parameter values are the same as in Fig. \ref{fig3}.}\label{fig4}
\end{figure}
\begin{figure}[htbp]
\centering
  \includegraphics[width=\linewidth]{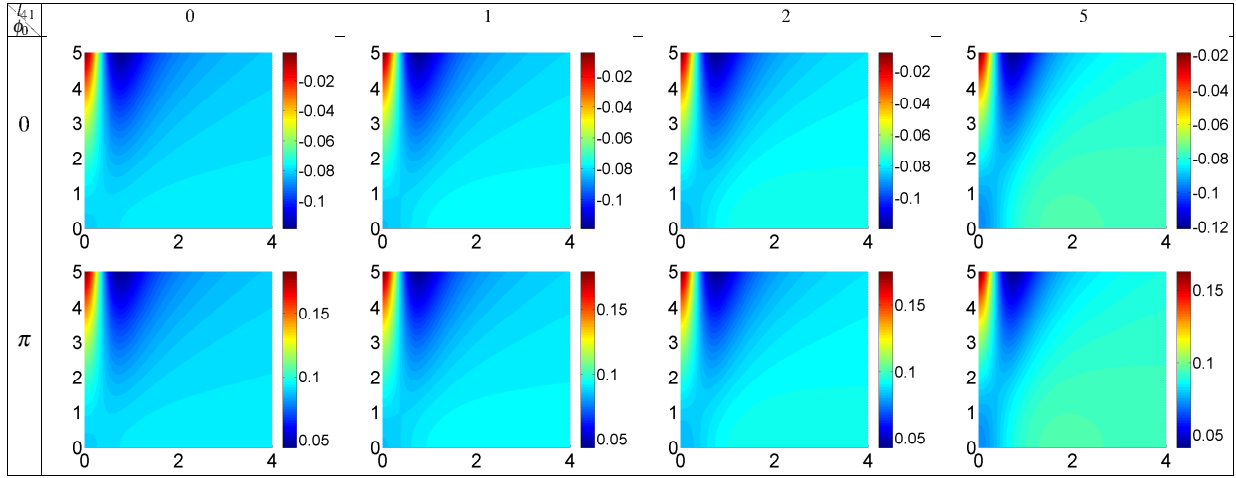}
  \caption{\small The corresponding $z$-component of Fig. \ref{fig4} with the same values of parameters.}\label{fig5}
\end{figure}

Figure \ref{fig4} demonstrates the relation of the dimensionless group velocity profile of the probe LG field with the dimensionless propagation distance $z/z_R$ (horizontal axis) and the radial coordinate $r/w_0$ (vertical axis). It should be pointed out that different modes of the probe LG field are employed in this figure, i.e. $l_{41}=0,1,2,5$, together with two different relative phases: $\phi_0=0$ (first row) and $\pi$ (second row). Moreover, all panels are obtained for zero probe detuning $\Delta_{41}=0$, whereas other parameter values are the same as in Fig. \ref{fig3}. It is obvious from Fig. \ref{fig3} that the group velocity should be less than $c$ at $r=z=0$ when the relative phase is chosen to be $\phi_0=0$. The first row of Fig. \ref{fig4} illustrates the subluminal group velocity of the probe LG field inside the atomic medium with $\phi_0=0$. It is also noticeable that the distribution of subluminality is concentrated at the smaller propagation distances for the higher-order LG modes. Furthermore, the probe LG field may propagate inside the atomic medium with the superluminal group velocity when the relative phase is shifted to $\phi_0=\pi$, and it is evident that the second row of Fig. \ref{fig4} provides evidence for this possibility. The different values of OAM results in the various behaviors for the superluminal group velocity spreading over the propagation direction, however, the variation between the group velocity profiles of different OAMs is not so apparent in the Rayleigh range.

Figure \ref{fig5} shows the corresponding $z$-component of Fig. \ref{fig4} with the same values of parameters. We figured out that there is a difference between the magnitude of the group velocity vector and that of $z$-component. An investigation of Figs. \ref{fig4} and \ref{fig5} indicates that the slightly difference between these two quantities refers to the behavior of subluminality and superluminality spreading over the propagation direction.
\begin{figure}[htbp]
\centering
  \includegraphics[width=0.75\linewidth]{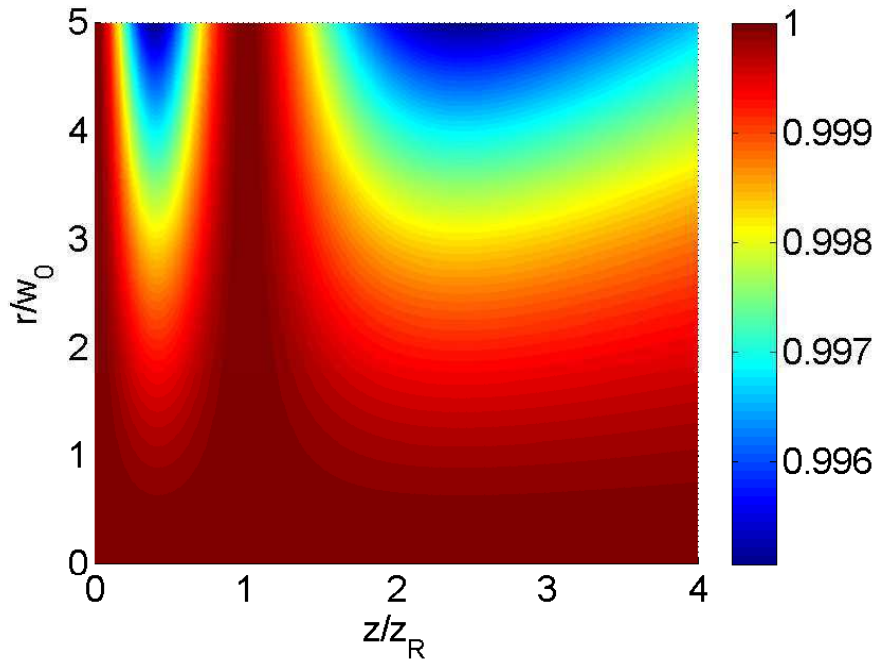}
  \caption{\small The coefficient of $v_g$ in Eq. (\ref{e10}) as a function of $z/z_R$ and $r/w_0$ for $l_{41}=0$. Other used parameters are the same as in Fig. \ref{fig3}.}\label{fig6}
\end{figure}

For the best understanding of this difference, the coefficient of $v_g$ in Eq. (\ref{e10}) is plotted as a function of $z/z_R$ and $r/w_0$ for $l_{41}=0$ in Fig. \ref{fig6}. It should be pointed out that other used parameters are the same as in Fig. \ref{fig3}. Since the coefficient has a negligible dependence on the OAM of the probe field ($l_{41}$), it is presented just for $l_{41}=0$. According to Fig. \ref{fig6}, the value of $v_g$ is different from that of $v_z$ everywhere except at the waist of the beam, the Rayleigh range, and on the optical axis; however, the most differences have appeared at the points away from the optical axis.

\subsection{The group velocity of the probe LG field in an atomic medium with a spatially dependent refractive index}

Finally, we prepare an atomic medium with a spatially dependent refractive index by considering one of the strong coupling fields as an LG field. Since the refractive index of the atomic medium depends on the real part of the susceptibility applying a strong coupling LG field to the atomic medium leads to its dependence on the space coordinates. As a result, the magnitude of the group velocity vector, which has three components, can not be expressed by a simple mathematical expression. Notice that due to the complexity and length of the resulting expression, it is not presented here. In this section, we would like to investigate the group velocity of the probe LG field at $z=0$. It is intriguing to note that the group velocity has only $z$-component, and the expression for its magnitude reads
\begin{eqnarray}\label{e11}
    v_g(r,\varphi)=\dfrac{c}{(n+\omega\partial n/\partial\omega)[1+\frac{2c^2}{n^2\omega^2w_0^2}(-\frac{r^2}{w_0^2}+(2p_{41}+|l_{41}|+1))]},
\end{eqnarray}
which can be reduced to Eq. (\ref{e7}) on the optical axis, $r=0$.

We proceed by regarding the strong coupling field of the $|2\rangle\leftrightarrow|4\rangle$ transition, $\Omega_{42}$, as an LG field with zero radial index, $p_{42}=0$, which takes the following form
\begin{eqnarray}\label{e12}
\Omega_{42}(r,\varphi)=\Omega_{42_0} (\frac{\sqrt{2}r}{w_{LG}})^{|l_{42}|}e^{-\frac{r^2}{w_{LG}^2}}e^{il_{42}\varphi}.
\end{eqnarray}
Here, the constant Rabi frequency, waist and OAM value of the strong coupling LG field are described by $\Omega_{42_0}$, $w_{LG}$, and $l_{42}$, respectively.
Consequently, the spatially dependent behavior of the medium absorption for zero probe detuning is given by
\begin{eqnarray}\label{e13}
Im(\chi)=\frac{-\gamma\Omega_{31}\Omega_{32}^\ast\Omega_{42}(r)\cos(l_{42}\varphi+\phi_0)}{A_1}.
\end{eqnarray}
Moreover, one can obtain the refractive index of the medium and its dispersion slope in the case of zero probe detuning which are respectively given by
\begin{eqnarray}\label{e14}
n=1+\frac{\gamma\Omega_{31}\Omega_{32}^\ast\Omega_{42}(r)\sin(l_{42}\varphi+\phi_0)}{2A_1},
\end{eqnarray}
and
\begin{eqnarray}\label{e15}
\frac{\partial n}{\partial\omega}=\frac{\Omega_{31}\Omega_{32}^\ast\Omega_{42}(r) A_2 \cos(l_{42}\varphi+\phi_0)}{2A_1 A_3}.
\end{eqnarray}
We now explore how the helical wavefront of the strong coupling LG field, which is indicated by its OAM, affects the response of the medium to the probe LG field [Eqs. (\ref{e13}, \ref{e15})], and its group velocity [Eq. (\ref{e11})] at $z=0$.
\begin{figure}[htbp]
\centering
  \includegraphics[width=\linewidth]{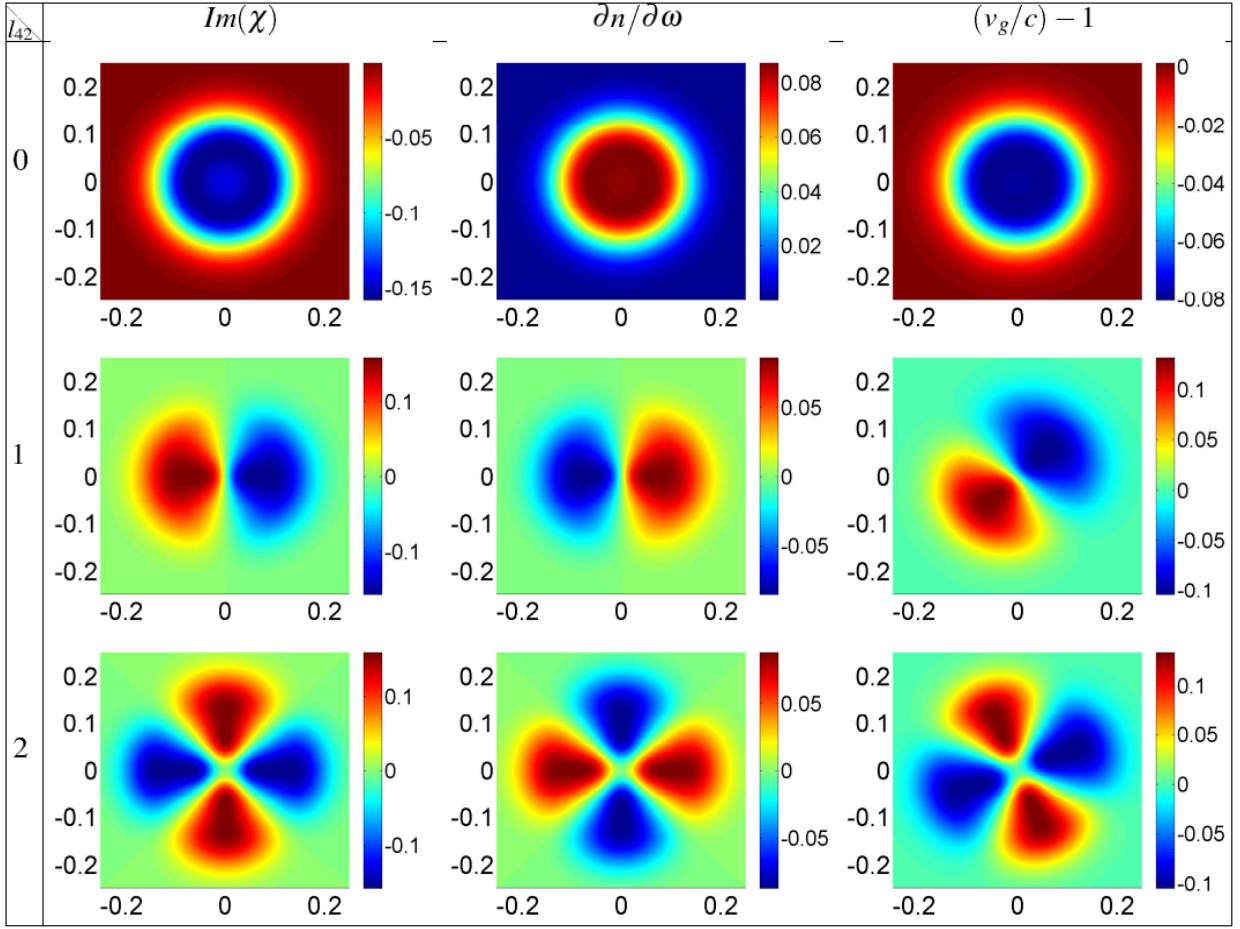}
  \caption{\small The imaginary part of the susceptibility (left column), dispersion slope (middle column), and dimensionless group velocity (right column) profiles as a function of $x$ (horizontal axis) and $y$ (vertical axis) for three different values of the OAM, i.e. $l_{42}=0,1,2$. In addition to the constant Rabi frequency of $\Omega_{42_0}=\gamma$, the size of the medium cross-section is fixed at $0.5 mm\times0.5 mm$ and the waist of the strong coupling LG field is considered to be $w_{LG}=100\mu m$. The waist and OAM value corresponding to the probe LG field are taken to be $w_0=100\mu m$ and $l_{41}=1$, respectively, whereas the other used parameters are the same as in Fig. \ref{fig3}.}\label{fig7}
\end{figure}

Figure \ref{fig7} illustrates the imaginary part of the susceptibility (left column), dispersion slope (middle column), and dimensionless group velocity (right column) profiles as a function of $x$ (horizontal axis) and $y$ (vertical axis) for three different values of the OAM, i.e. $l_{42}=0,1,2$. In addition to the constant Rabi frequency of $\Omega_{42_0}=\gamma$, we have fixed the size of the medium cross-section at $0.5 mm\times0.5 mm$ and the waist of the strong coupling LG field at $w_{LG}=100\mu m$. The waist and OAM value corresponding to the probe LG field are taken to be $w_0=100\mu m$ and $l_{41}=1$, respectively, whereas the other used parameters are the same as in Fig. \ref{fig3}. It is worth noting that the increase in the OAM value of probe LG field cannot have a considerable impact on the pattern of group velocity, while the OAM value of the strong coupling LG field can significantly change the behavior of group velocity. On the other hand, the first row of Fig. \ref{fig7} shows that the medium has a Gaussian-like gain response to the probe LG field with a normal dispersion and subluminal group velocity when $\Omega_{42}$ has a Gaussian function, $l_{42}=0$. In the second row of Fig. \ref{fig7}, a petal-like pattern has appeared whose right (left) petal displays a gain (absorption) region with normal (anomalous) dispersion, however, the group velocity profile behaves in an unusual manner. Although the subluminal (superluminal) group velocity exists in a medium characterized by normal (anomalous) dispersion, there is a region of normal (anomalous) dispersion in which the probe field experiences the superluminal (subluminal) group velocity. It should be mentioned that the main reason for this strange behavior lies in the counterclockwise rotation of the petal-like pattern related to the group velocity. As exhibited in the third row of Fig. \ref{fig7}, the number of petals and regions with unusual group velocity behavior has been doubled by $l_{42}=2$. Generally, one can conclude that the number of petals equals $2l_{42}$ in all panels.
\begin{figure}[htbp]
\centering
  \includegraphics[width=\linewidth]{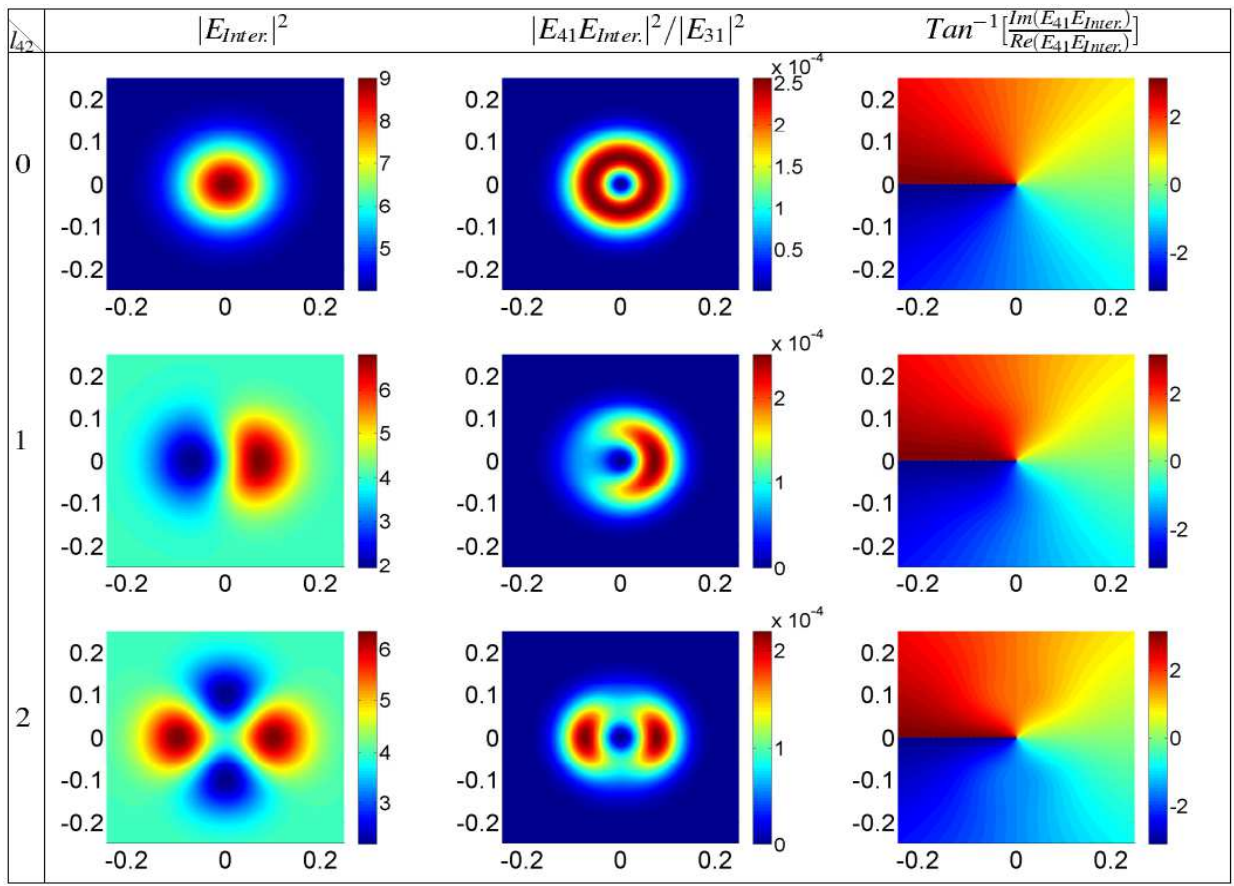}
  \caption{\small The interference pattern of the coupling fields (left column), intensity (middle column) and phase (right column) profiles of the probe LG field passing through the interference pattern as a function of $x$ and $y$ for three different values of the OAM, i.e. $l_{42}=0,1,2$. The constant amplitudes of the coupling and probe fields are considered to be $E_{31}=E_{32}=E_{42}$ and $E_{41}=0.01E_{31}$, respectively. Other parameters of the coupling and probe LG fields are the same as in Fig. \ref{fig7}.}\label{fig8}
\end{figure}

Let us proceed with the interference perspective to find a physical interpretation for the obtained results in Fig. \ref{fig7}. It is to be noted that two coupling planar fields ($E_{31}$ and $E_{32}$) and an LG coupling field ($E_{42}$) experience an interference inside the atomic medium. The spatially dependent interference pattern resulting from the helical phase front of the coupling LG field can change the properties of a probe LG field propagating through this pattern. Figure \ref{fig8} is plotted in order to study the mentioned changes. In this figure, the interference pattern of the three coupling fields (left column), intensity (middle column) and phase (right column) profiles of the probe LG field passing through the interference pattern are plotted as a function of $x$ and $y$ for three different values of the OAM, i.e. $l_{42}=0,1,2$. The constant amplitudes of the coupling and probe fields are considered to be $E_{31}=E_{32}=E_{42}$ and $E_{41}=0.01E_{31}$, respectively. Other parameters of the coupling and probe LG fields are the same as in Fig. \ref{fig7}. The comparison between the left columns of Figs. \ref{fig7} and \ref{fig8} illustrates that the gain (absorption) regions originate from the constructive (destructive) interference of the three coupling fields in the atomic medium. In the middle column of Fig. \ref{fig8}, it can easily be seen that the probe LG field propagating through the spatially dependent interference pattern is amplified (attenuated) in the constructive (destructive) interference regions. The first row of the right column shows that the phase front of the probe LG field has an undisturbed helical shape considering the Gaussian mode for $E_{42}$. It is interesting to note that the probe LG field undergoes a distortion in its phase front as $E_{42}$ has the helical phase front with nonzero OAM value ($l_{42}=1$ or $2$). It should be pointed out that the evolution of the probe LG field can be investigated using the susceptibility of the medium, calculated through the density matrix formalism, and wave propagation equation ($\partial E_{41}(z)/\partial z=ik_p\chi E_{41}(z)/2$) at $L=0.5\mu m$. Employing this manner yields the same results as those of the middle and right columns of Fig. \ref{fig8} for the intensity and phase profiles of the probe LG field ($E_{41}(z=L)$). It means that the strange behavior related to the group velocity of the probe LG field inside the dispersive medium lies in the distortion of the helical phase front of the LG probe field due to the classical interference pattern of the coupling planar and LG fields.
\begin{figure}[htbp]
\centering
  \includegraphics[width=\linewidth]{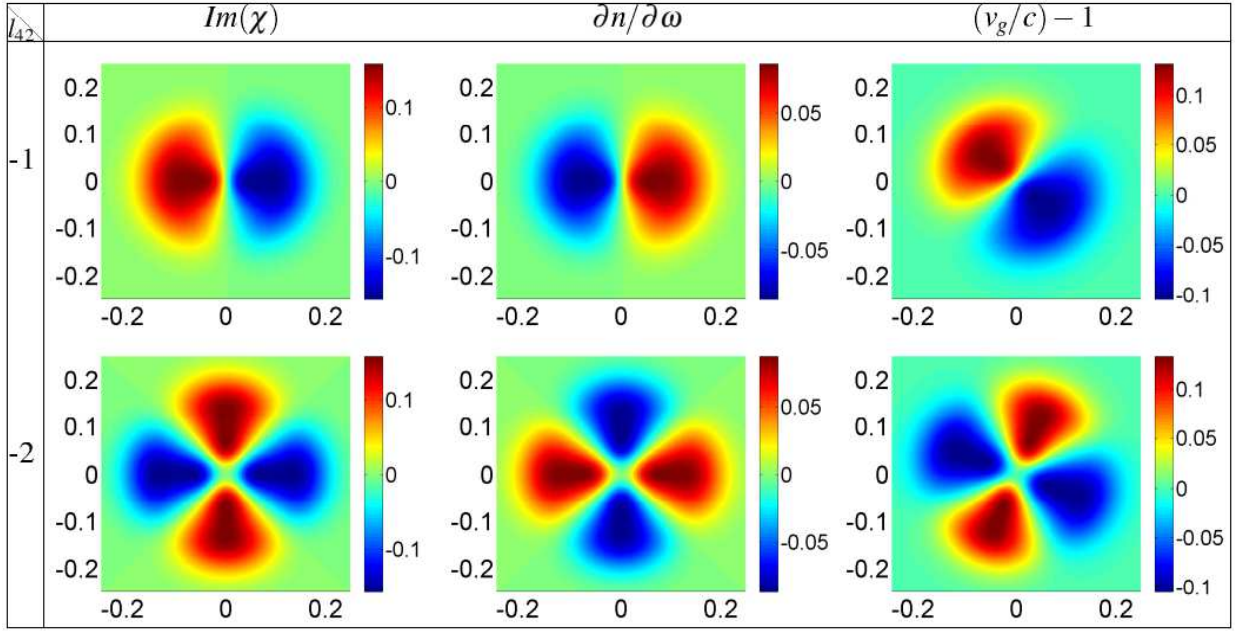}
  \caption{\small The imaginary part of the susceptibility (left column), dispersion slope (middle column), and dimensionless group velocity (right column) profiles versus $x$ and $y$ for two negative values of the OAM, i.e. $l_{42}=-1,-2$. The values of parameters are the same as in Fig. \ref{fig7}.}\label{fig9}
\end{figure}

In Fig. \ref{fig9}, we replot six panels of Fig. \ref{fig7} for the strong coupling LG field with reverse helical wavefront, i.e. the negative value of the OAM. According to Eqs. (\ref{e13}) and (\ref{e15}), it is expected that the imaginary part of the susceptibility and refractive index of the atomic medium do not change for the negative values of the OAM. However, the group velocity has been affected by negative OAMs, and its petals rotate clockwise, which results in the change of the position of regions with unusual behaviors.

\begin{figure}[htbp]
\centering
  \includegraphics[width=\linewidth]{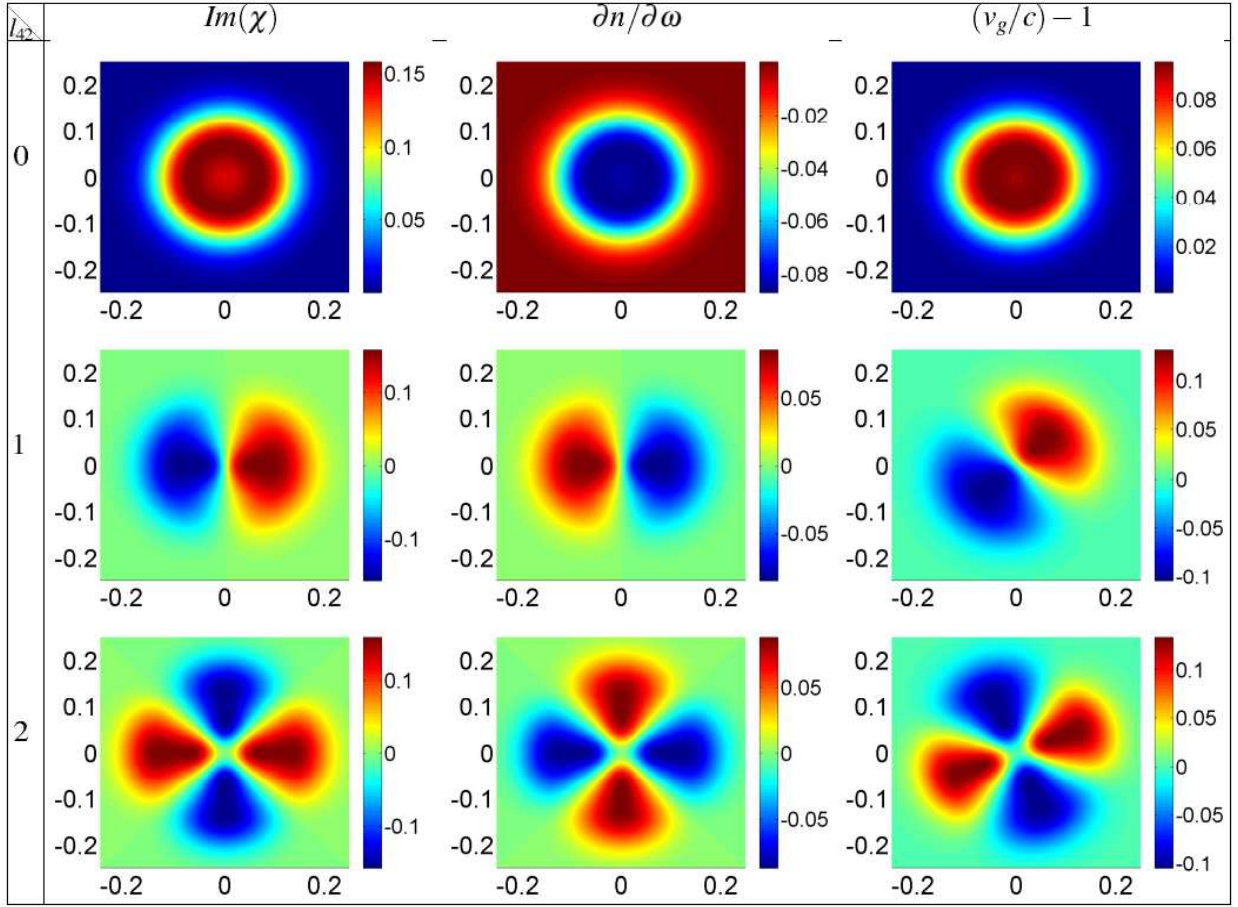}
  \caption{\small The imaginary part of the susceptibility (left column), dispersion slope (middle column), and dimensionless group velocity (right column) profiles as a function of $x$ and $y$ for three different values of the OAM, i.e. $l_{42}=0,1,2$, and $\phi_0=\pi$ with the same parameters of Fig. \ref{fig7}.}\label{fig10}
\end{figure}

Figure \ref{fig10} is just the same as Fig. \ref{fig7}, but for the shift of the relative phase from $0$ to $\pi$. It is apparent from Eqs. (\ref{e13}) and (\ref{e15}) that the absorption and anomalous dispersion regions turn into the gain and normal dispersion ones, respectively, and vise versa. Moreover, it is shown that a simple change in the relative phase of applied fields leads to the interchange of the regions with unusual behavior of group velocity. However, the rotation direction of the petal-like patterns of the group velocity remains counterclockwise, which means that it is directly dependent on the sign of the OAM of the strong coupling LG field.

In summary, we have investigated the group velocity of the LG beam in a four-level double $V$-type atomic system. We have derived an analytical expression for the group velocity of the probe LG field on its optical axis through the uniformly dispersive medium. This result obtained with the aid of the well-known formula of the group velocity, $v_g=|\nabla\partial_\omega\Phi|^{-1}$, is consistent with the group velocity of an LG beam in free space. Moreover, the behavior of the group velocity corresponding to the probe LG field and its $z$-component along the propagation direction has been analytically studied and shown that they both are of equal value on the optical axis, at the waist of the beam and the Rayleigh range.  In another scenario, we have demonstrated how the OAM of another LG field can affect the group velocity of the probe LG field. It is interesting to note that the superluminal group velocity may exist in the normal dispersion regions of a spatially dependent dispersive medium, while the subluminal group velocity can be found in the anomalous dispersion regions simultaneously. We have proved that this strange behavior is physically rooted in the classical interference of two coupling planar and an LG fields inside the atomic medium, which leads to the distortion in the phase front of the probe LG field.

\section*{Acknowledgements}
Z.A.S. acknowledges financial support from Iran's National Elites Foundation (INEF) (Shahid Chamran's Scientific Prize, Grant No. 15/10597).


\end{document}